\documentclass[12pt]{article}
\usepackage[latin1]{inputenc}
\usepackage[british]{babel}
\usepackage{cmap}
\usepackage{lmodern}

\usepackage{amssymb, amsmath, amsthm}
\usepackage[a4paper,top=25mm,bottom=25mm,left=25mm,right=25mm]{geometry}
\usepackage{etex}
\usepackage{ragged2e}

\usepackage{authblk} % for headings
\usepackage{pifont}
\usepackage{graphicx}
\usepackage[usenames,dvipsnames,svgnames,table]{xcolor}
\usepackage[figuresright]{rotating}
\usepackage{xtab} % tackle the long tables
\usepackage{longtable} % tackle the long tables
\usepackage{ltxtable} % longtable with tabularx
\usepackage{multirow}
\usepackage{footnote}
\usepackage[stable]{footmisc}
\usepackage{chngpage} % allows for temporary adjustment of side margins
\usepackage{pdflscape} % landscape environment
\usepackage[nottoc,notlot,notlof]{tocbibind} % includes everything in ToC

\usepackage{pgfplots}
\pgfplotsset{compat=1.14}
\pgfplotsset{every tick label/.append style={font=\footnotesize}}
\usepackage{setspace}

\usepackage{array}
\newcolumntype{K}[1]{>{\centering\arraybackslash$}p{#1}<{$}}
\usepackage{wasysym}

\makesavenoteenv{tabular}
\usepackage{tabularx}
\usepackage{booktabs}
\usepackage[flushleft]{threeparttable}
\usepackage[referable]{threeparttablex} % footnotes in tabu
\newcolumntype{R}{>{\raggedleft\arraybackslash}X}
\newcolumntype{L}{>{\raggedright\arraybackslash}X}
\newcolumntype{C}{>{\centering\arraybackslash}X}
\newcolumntype{M}[1]{>{\centering\arraybackslash}m{#1}}
\newcolumntype{A}{>{\columncolor{gray!25}}C}
\newcolumntype{a}{>{\columncolor{gray!25}}c}

\newlength{\tablen}

\usepackage{dcolumn} % alignment to decimal points
\newcolumntype{.}{D{.}{.}{-1}}

\usepackage{tikz}
\usetikzlibrary{arrows, automata, calc, matrix, patterns, positioning, trees}
\usepackage[semicolon]{natbib}
\usepackage[hyphens]{url}
\makeatletter
\g@addto@macro{\UrlBreaks}{\UrlOrds}
\makeatother
\usepackage{hyperref} % [hidelinks]
\hypersetup{
  colorlinks   = true,    		% Colours links instead of ugly boxes
  urlcolor     = blue,    		% Colour for external hyperlinks
  linkcolor    = blue,    		% Colour of internal links
  citecolor    = ForestGreen  	% Colour of citations
}
\usepackage{microtype}
\usepackage[justification=centerfirst]{caption} % [justification=centering]

\usepackage[labelformat=simple]{subcaption}

\DeclareCaptionLabelFormat{parenthesis}{(#2)}
\makeatletter
\renewcommand\p@subfigure{\arabic{figure}.}
\makeatother

\DeclareCaptionLabelFormat{parenthesis}{(#2)}
\makeatletter
\renewcommand\p@subtable{\arabic{table}.}
\makeatother

\def\addlegendimage{\csname pgfplots@addlegendimage\endcsname}

\usepackage{enumitem}
\setlist[itemize]{leftmargin=2.5\parindent}
\setlist[enumerate]{leftmargin=2.5\parindent}

\theoremstyle{plain}
\theoremstyle{definition}
\theoremstyle{remark}

\newcommand{\down}{\textcolor{BrickRed}{\rotatebox[origin=c]{270}{\ding{212}}}}
\newcommand{\up}{\textcolor{PineGreen}{\rotatebox[origin=c]{90}{\ding{212}}}}

\def\keywords{\vspace{.5em} % Add keywords
{\noindent \textit{Keywords}: }}

\def\JEL{\vspace{.5em} % Add keywords
{\noindent \textbf{\emph{JEL} classification number}: }}

\def\AMS{\vspace{.5em} % Add keywords
{\noindent \textbf{\emph{MSC} class}: }}

\author{\href{https://sites.google.com/view/laszlocsato}{L\'aszl\'o Csat\'o}\thanks{~E-mail: \emph{laszlo.csato@sztaki.hu}} }
\affil{Institute for Computer Science and Control (SZTAKI) \\
Laboratory on Engineering and Management Intelligence \\ Research Group of Operations Research and Decision Systems}
\affil{Corvinus University of Budapest (BCE) \\
Department of Operations Research and Actuarial Sciences}
\affil{Budapest, Hungary}

\title{The UEFA Champions League seeding is not strategy-proof since the 2015/16 season}
\date{\today}

\def\Dedication{
{\noindent
$\mathfrak{Hat}$ $\mathfrak{die}$ $\mathfrak{Seele}$ $\mathfrak{einmal}$ $\mathfrak{eine}$ $\mathfrak{bestimmte}$ $\mathfrak{Richtung}$ 
$\mathfrak{fort}$ $\mathfrak{zum}$ $\mathfrak{Ziele}$ $\mathfrak{oder}$ $\mathfrak{zur\ddot{u}ckgewendet}$ $\mathfrak{nach}$ $\mathfrak{einem}$ $\mathfrak{Rettungshafen}$, $\mathfrak{so}$ $\mathfrak{geschieht}$ $\mathfrak{es}$ $\mathfrak{leicht}$, $\mathfrak{da \ss}$ $\mathfrak{die}$ $\mathfrak{Gr\ddot{u}nde}$, $\mathfrak{welche}$ $\mathfrak{den}$ $\mathfrak{einen}$ $\mathfrak{zum}$ \linebreak $\mathfrak{Innehalten}$ $\mathfrak{n\ddot{o}tigen}$, $\mathfrak{den}$ $\mathfrak{anderen}$ $\mathfrak{zum}$ $\mathfrak{Unternehmen}$ $\mathfrak{berechtigen}$, $\mathfrak{nicht}$ $\mathfrak{leicht}$ $\mathfrak{in}$ $\mathfrak{ihrer}$ $\mathfrak{ganzen}$ $\mathfrak{St\ddot{a}rke}$ $\mathfrak{gef\ddot{u}hlt}$ $\mathfrak{werden}$, $\mathfrak{und}$ $\mathfrak{da}$ $\mathfrak{die}$ $\mathfrak{Handlung}$ $\mathfrak{indes}$ $\mathfrak{fortschreitet}$, $\mathfrak{so}$ $\mathfrak{kommt}$ $\mathfrak{man}$ $\mathfrak{im}$ $\mathfrak{Strom}$ $\mathfrak{der}$ $\mathfrak{Bewegung}$ $\mathfrak{\ddot{u}ber}$ $\mathfrak{die}$ $\mathfrak{Grenze}$ $\mathfrak{des}$ $\mathfrak{Gleichgewichts}$, $\mathfrak{\ddot{u}ber}$ $\mathfrak{die}$ $\mathfrak{Kulminationslinie}$ $\mathfrak{hinaus}$, $\mathfrak{ohne}$ $\mathfrak{es}$ $\mathfrak{gewahr}$ $\mathfrak{zu}$ $\mathfrak{werden}$.\footnote{~
``\emph{When once the mind has taken a decided direction towards an object, or turned back towards a harbour of refuge, it may easily happen that the motives which in the one base naturally serve to restrain, and those which in the other as naturally excite to enterprise, are not felt at once in their full force; and as the progress of action in the mean time continues, one is carried along by the stream of movement beyond the line of equilibrium, beyond the culminating point, without being aware of it.}'' (Source: Carl von Clausewitz: \emph{On War}, Book 7, Chapter 21---Invasion. Translated by Colonel James John Graham, London, N. Tr\"ubner, 1873. \url{http://clausewitz.com/readings/OnWar1873/TOC.htm})}
}
\vspace{0.25cm}

\flushright
\noindent (Carl von Clausewitz: \emph{Vom Kriege})

\vspace{1cm} 
\justify }

\begin{document}

\newgeometry{top=10mm,bottom=10mm,left=25mm,right=25mm}
\maketitle
\thispagestyle{empty}
\Dedication

\begin{abstract}
\noindent
% ORL limit: 80 words
%The current seeding regime of the UEFA (Union of European Football Associations) Champions League is shown to be incentive incompatible since the 2015/16 season. The reason is giving the slot of the titleholder in the first pot to the champion of a lower-ranked association if the titleholder is the champion of a high-ranked league. Filling all vacancies through the national leagues excludes the presence of perverse incentives. UEFA is encouraged to introduce this policy from the 2021-24 cycle onwards.
Fairness has several interpretations in sports, one of them being that the rules should guarantee incentive compatibility, namely, a team cannot be worse off due to better results in any feasible scenario. The current seeding regime of the most prestigious annual European club football tournament, the UEFA (Union of European Football Associations) Champions League, is shown to violate this requirement since the 2015/16 season. In particular, if the titleholder qualifies for the first pot by being a champion in a high-ranked league, its slot is given to a team from a lower-ranked association, which can harm a top club from the domestic championship of the titleholder. However, filling all vacancies through the national leagues excludes the presence of perverse incentives. UEFA is encouraged to introduce this policy from the 2021-24 cycle onwards.


\AMS{62F07, 91B14}
% Ranking and selection
% Social choice

\JEL{C44, D71, Z20}
% Operations Research, Statistical Decision Theory 
% Social Choice, Clubs, Committees, Associations
% Sports Economics, General
\end{abstract}

\clearpage
\restoregeometry

\section{Introduction} \label{Sec1}

It is widely accepted that every sporting contest should provide the appropriate incentives to perform \citep{Szymanski2003}. However, this simple requirement does not always hold as several historical cases attest \citep{PrestonSzymanski2003, KendallLenten2017}.
A classical example is presented by designs where the lowest-ranked team receives the first draft pick in the following season, which makes losing profitable after a team is eliminated from the later rounds (e.g.\ play-offs) of the tournament \citep{TaylorTrogdon2002, BalsdonFongThayer2007}. Since managers are found to apply concrete tanking strategies \citep{Fornwagner2019}, it would be important to adopt a policy that ensures integrity \citep{Lenten2016, LentenSmithBoys2018, BanchioMunro2020}.

According to recent game-theoretical studies \citep{Pauly2014, Vong2017}, incentive incompatibility sometimes cannot be avoided because the unique theoretical solution would be too harsh to implement in practice, for example, by allowing only the top team to proceed from a round-robin tournament.
On the other hand, in certain cases, there exists an (almost) costless guarantee of fairness, and there is even some development towards this outcome in the real-world.

To mention some illustrative cases, \citet{DuranGuajardoSaure2017} demonstrate the openness of the governing bodies in football to improve fairness by rescheduling the FIFA World Cup South American qualifiers. The drawing procedure of the 2018 FIFA World Cup was reformed to resemble one of the suggestions in \citet{Guyon2015a}. UEFA used the results of \citet{Guyon2018a} to modify the knockout bracket in the UEFA European Championship 2020 to minimise group advantage.

There is also an evolution in the direction of incentive compatibility. For instance, tournament systems, consisting of one round-robin and multiple knockout tournaments with noncumulative prizes, are proved to satisfy strategy-proofness only if all vacant slots are awarded based on the results of the round-robin tournament \citep{DagaevSonin2018}. The qualification to the two annual European club football tournaments organised by the UEFA (Union of European Football Associations) was incentive incompatible due to this result: the entry of the more prestigious UEFA Champions League between 2015 and 2018 \citep{Csato2019b}, and the entry of the second-tier competition UEFA Europa League before 2016 \citep{DagaevSonin2018}. However, these mistakes have been corrected, and now no team can be strictly better off by losing in both championships.

In the following, we will present that the seeding regime applied in the group stage of the UEFA Champions League from the 2015/16 season leads to another form of incentive incompatibility: the rules may punish a team for better results in its domestic championship by seeding it in a weaker pot. Naturally, a straightforward solution is also provided.

This is probably the first paper analysing the draw systems of sports tournaments with respect to strategy-proofness, which is our main contribution.
On the other hand, the effects of the seeding reforms in the UEFA Champions League have been recently evaluated via Monte-Carlo simulations in \citet{DagaevRudyak2019} and \citet{CoronaForrestTenaWiper2019}.
The current article is strongly connected to the works investigating the draw of round-robin groups under some geographical and/or seeding restrictions, too \citep{Guyon2015a, LalienaLopez2019, CeaDuranGuajardoSureSiebertZamorano2020}. The procedure of the UEFA Champions League Round of 16 draw has been considered by \citet{KlossnerBecker2013}, as well as by \citet{BoczonWilson2018}.
The importance of our analysis is reinforced by the fact that the draws of the UEFA Champions League are regularly discussed in the mainstream media \citep{Guyon2015b, Guyon2017c, Guyon2017b, Guyon2017a, Guyon2018c, Guyon2019e, Guyon2019g, Guyon2019c, Guyon2019f}.

Unfortunately, although badly designed tournaments may have an adverse effect on efforts and fairness, \citet{HaugenKrumer2019} reveal that the sport management literature has largely ignored this issue in recent years. Hopefully, our research will contribute to call the attention of sports administrators to the importance of tournament design.

%\citet{Guyon2019b} proposes a new knockout format for the Round of 16 of the Champions League.

The structure of the paper is as follows.
Section~\ref{Sec2} presents a hypothetical example with a slight modification of real-world match results to motivate our approach. The consequences of the problem are discussed in Section~\ref{Sec3}. Section~\ref{Sec4} contains our proposal for guaranteeing incentive compatibility. Finally, the main message is summarised in Section~\ref{Sec5}.

\section{An illustrative example} \label{Sec2}

The participants of the \href{https://en.wikipedia.org/wiki/2015\%E2\%80\%9316_UEFA_Champions_League}{2015/16 UEFA Champions League} were determined by the previous season of the national leagues across the continent, as well as by the two European club competitions, the UEFA Champions League (shortly Champions League or simply CL) and the UEFA Europa League (shortly Europa League or simply EL).
For what follows, assume the following counterfactual modifications to realized results:
\begin{itemize}
\item
Sevilla FC defeated FC Barcelona in the Spanish La Liga on 11 April 2015 (the real result was 2-2);
\item
SK Rapid Wien advanced to the Champions League group stage from the play-off round of the League Route in the qualifying against FC Shakhtar Donetsk (in fact, FC Shakhtar Donetsk won 3-2 on aggregate).
\end{itemize}
In this case, the Spanish national league would have been won by Real Madrid CF as it would have 92 points similarly to FC Barcelona but better head-to-head results, which was the first tie-breaking criterion (the outcomes of the clashes Real Madrid CF vs.\ FC Barcelona were 3-1 in Madrid and 1-2 in Barcelona, see \url{https://en.wikipedia.org/wiki/2014\%E2\%80\%9315_La_Liga}).
Nonetheless, the CL titleholder, FC Barcelona, would have also qualified through its domestic championship, thus the vacant slot would have been filled by the EL titleholder Sevilla FC from Spain, despite finishing only fifth in La Liga \citep[Article~3.04]{UEFA2015a}.

The draw of the Champions League group stage was regulated by \citet[Article~13.05]{UEFA2015a} as follows: \\
``\emph{For the purpose of the draw, the 32 clubs involved in the group stage are seeded into four groups of eight. The first group comprises the titleholder (top seed) and the domestic champions of the seven top-ranked associations in accordance with the access list (see Annex A). If the titleholder is one of the top seven associations' domestic champions, the  group is completed with the champion of the association ranked eight. The other three groups are composed in accordance with the club coefficient rankings established at the  beginning of the season (see Annex D).}''

\begin{table}[!ht]
  \centering
  \caption{Pot composition in the (hypothetical) 2015/16 UEFA Champions League if Sevilla FC lost both matches against Real Madrid CF in the Spanish league \\\vspace{0.25cm}
  \footnotesize{Arrows indicate changes in the pots if Sevilla FC would have scored at least one point against Real Madrid CF in 2014/15 Spanish La Liga. \\
Teams written in \textbf{bold} qualified directly for the group stage. Coefficient stands for the UEFA club coefficient. Source: \url{https://www.footballseeding.com/club-ranking/a2014-2015/}.}
  }
  \label{Table1}
\footnotesize
\begin{subtable}{\textwidth}
  \centering
  \caption{Pot 1}
  \label{Table1a}
\rowcolors{1}{}{gray!20}
    \begin{tabularx}{\textwidth}{Lc Lr} \toprule \hiderowcolors
    Club  & & Association (position) & Coefficient \\ \bottomrule \showrowcolors
    \textbf{Real Madrid CF} & \down \textcolor{gray!20}{\rotatebox[origin=c]{270}{\ding{212}}} & Spain (champion) & 171.999 \\
    \textbf{FC Barcelona} & & Spain (runner-up, CL titleholder) & 164.999 \\
    \textbf{Chelsea FC} & & England (champion) & 142.078 \\
    \textbf{FC Bayern M\"unchen} & & Germany (champion) & 154.833 \\
    \textbf{Juventus} & & Italy (champion) & 95.102 \\
    \textbf{SL Benfica} & & Portugal (champion) & 118.276 \\
    \textbf{Paris Saint-German FC} & & France (champion) & 100.483 \\
    \textbf{FC Zenit St Petersburg} & & Russia (champion) & 90.099 \\ \bottomrule
    \end{tabularx}
\end{subtable}

\vspace{0.25cm}
\begin{subtable}{\textwidth}
  \centering
  \caption{Pot 2}
  \label{Table1b}
\rowcolors{1}{}{gray!20}
    \begin{tabularx}{\textwidth}{Lc Lr} \toprule \hiderowcolors
    Club  & & Association (position) & Coefficient \\ \bottomrule \showrowcolors
    \textbf{Club Atl\'etico de Madrid} & & Spain (3rd) & 120.999 \\
    \textbf{FC Porto} & & Portugal (runner-up) & 111.276 \\
    \textbf{Arsenal} & & England (3rd) & 110.078 \\
    Manchester United FC & & England (4th) & 103.078 \\
    Valencia CF & & Spain (4th) & 99.999 \\
    Bayer 04 Leverkusen & & Germany (4th) & 87.883 \\
    \textbf{Manchester City FC} & & England (runner-up) & 87.078 \\
    \textbf{Sevilla FC} & \down \textcolor{white}{\rotatebox[origin=c]{270}{\ding{212}}} & Spain (5th, EL titleholder) & 80.499 \\ \bottomrule
    \end{tabularx}
\end{subtable}

\vspace{0.25cm}
\begin{subtable}{\textwidth}
  \centering
  \caption{Pot 3}
  \label{Table1c}
\rowcolors{1}{}{gray!20}
    \begin{tabularx}{\textwidth}{Lc Lr} \toprule \hiderowcolors
    Club  & & Association (position) & Coefficient \\ \bottomrule \showrowcolors
    \textbf{Olympique Lyonnais} & & France (runner-up) & 72.983 \\
    \textbf{FC Dinamo Kyiv} & & Ukraine (champion) & 65.033 \\
    \textbf{Olympiacos FC} & & Greece (champion) & 62.380 \\
    \textbf{PSV Eindhoven} & \up \up & Netherlands (champion) & 58.195 \\
    PFC CSKA Moskva & & Russia (runner-up) & 55.599 \\
    \textbf{Galatasaray A\c{S}} & & Turkey (champion) & 50.020 \\
    \textbf{AS Roma} & & Italy (runner-up) & 43.602 \\
    FC BATE Borisov & & Belarus (champion) & 35.150 \\ \bottomrule
    \end{tabularx}
\end{subtable}

\vspace{0.25cm}
\begin{subtable}{\textwidth}
  \centering
  \caption{Pot 4}
  \label{Table1d}
\rowcolors{1}{}{gray!20}
    \begin{tabularx}{\textwidth}{Lc Lr} \toprule \hiderowcolors
    Club  & & Association (position) & Coefficient \\ \bottomrule \showrowcolors
    \textbf{VfL Borussia M\"onchengladbach} & & Germany (3rd) & 33.883 \\
    \textbf{VfL Wolfsburg} & \textcolor{white}{\rotatebox[origin=c]{270}{\ding{212}}} \textcolor{white}{\rotatebox[origin=c]{270}{\ding{212}}} & Germany (runner-up) & 31.883 \\
    GNK Dinamo Zagreb & & Croatia (champion) & 24.700 \\
    Maccabi Tel-Aviv FC & & Israel (champion) & 18.200 \\
    SK Rapid Wien & & Austria (runner-up) & 15.635 \\
    \textbf{KAA Gent} & & Belgium (champion) & 13.440 \\
    Malm\"o FF & & Sweden (champion) & 12.545 \\
    FC Astana & & Kazakhstan (champion) & 3.825 \\ \bottomrule
    \end{tabularx}
\end{subtable}
\end{table}

Table~\ref{Table1} shows the composition of the pots in the scenario above.
Note that the CL titleholder is not a domestic champion of one of the top seven associations, and Sevilla FC is the lowest-ranked team of Pot 2.

Consider what happens if Sevilla FC would have scored at least one point against Real Madrid CF in the \href{https://en.wikipedia.org/wiki/2014\%E2\%80\%9315_La_Liga}{2014/15 season of the Spanish national league}---the real results of the matches Sevilla FC vs.\ Real Madrid CF were 2-3 in Sevilla and 1-2 in Madrid. Then FC Barcelona would have won La Liga and the teams in the Champions League would have remained the same.
However, the composition of the pots would have changed as indicated by the arrows in Table~\ref{Table1}:
\begin{itemize}
\item
PSV Eindhoven from the Netherlands would have been in Pot 1 due to being the champion of the association ranked eight;
\item
Real Madrid CF would have been relegated to Pot 2 due to not being the champion in Spain;
\item
Sevilla FC would have been relegated to Pot 3 due to having the lowest UEFA club coefficient in Pot 2 of our hypothetical scenario.
\end{itemize}
%Pot 1 contains the champion of the association ranked eight, PSV Eindhoven from the Netherlands, instead of Real Madrid CF. Hence the latter club is drawn from Pot 2, while Sevilla FC is relegated to Pot 3.

According to this outlined (hypothetical) scenario, the seeding rules of the Champions League group stage can punish Sevilla FC for having more favourable results in its domestic championship, as it would face a team from the stronger Pot 2 instead of the weaker Pot 3.
Note that the composition of the seeding pots does not depend on whether a team qualifies as a CL/EL titleholder, or directly through its domestic championship. 

\section{Discussion} \label{Sec3}

What is the cost of being seeded in Pot 3 rather than in Pot 2?
\citet{CoronaForrestTenaWiper2019} analyse the effects of the new seeding regime for the teams participating in the 2015/16 Champions League. Compared to the original seeding based exclusively on the UEFA club coefficients, FC Shakhtar Donetsk was lowered to Pot 3, which lead to a substantial reduction in the probability of qualifying to the first knockout round, from $0.633$ to $0.483$.

We have also attempted to quantify how Sevilla FC suffers from the unfair rule. For this purpose, the club Elo ratings from \url{http://clubelo.com/} have been used. It can be a better measure of current abilities than the UEFA club coefficient: the latter does not consider the results in the domestic league and is relatively inert due to being an average over the last five seasons. 

Elo rating quantifies the strength of each club on the basis of its past results such that winning against a stronger team is more valuable, while the influence of a game decreases when new matches are played. In contrast to the UEFA club coefficient, club Elo also reflects home advantage and goal difference.
Elo-inspired methods provide good predictive performance \citep{LasekSzlavikBhulai2013}, and have been extensively applied in the scientific literature \citep{HvattumArntzen2010, LasekSzlavikGagolewskiBhulai2016, CeaDuranGuajardoSureSiebertZamorano2020, Csato2020b}. In particular, \citet{Csato2020h} uses the same dataset to illuminate the impact of reforming the Champions League qualification in 2018.

In order to take into account the dynamic nature of this estimation of strength, the average of the Elo ratings on the day of the group stage draw (27 August 2015, see \url{http://clubelo.com/2015-08-27/Data}) and one day after the last match of the group stage (10 December 2015, \url{http://clubelo.com/2015-12-10/Data}) have been considered.
Furthermore, clubs from the same association could not be drawn against each other in the group stage of the Champions League, therefore Sevilla FC is not allowed to play against Real Madrid, Club Atl\'etico de Madrid, and Valencia CF if it would be drawn from Pot 3.

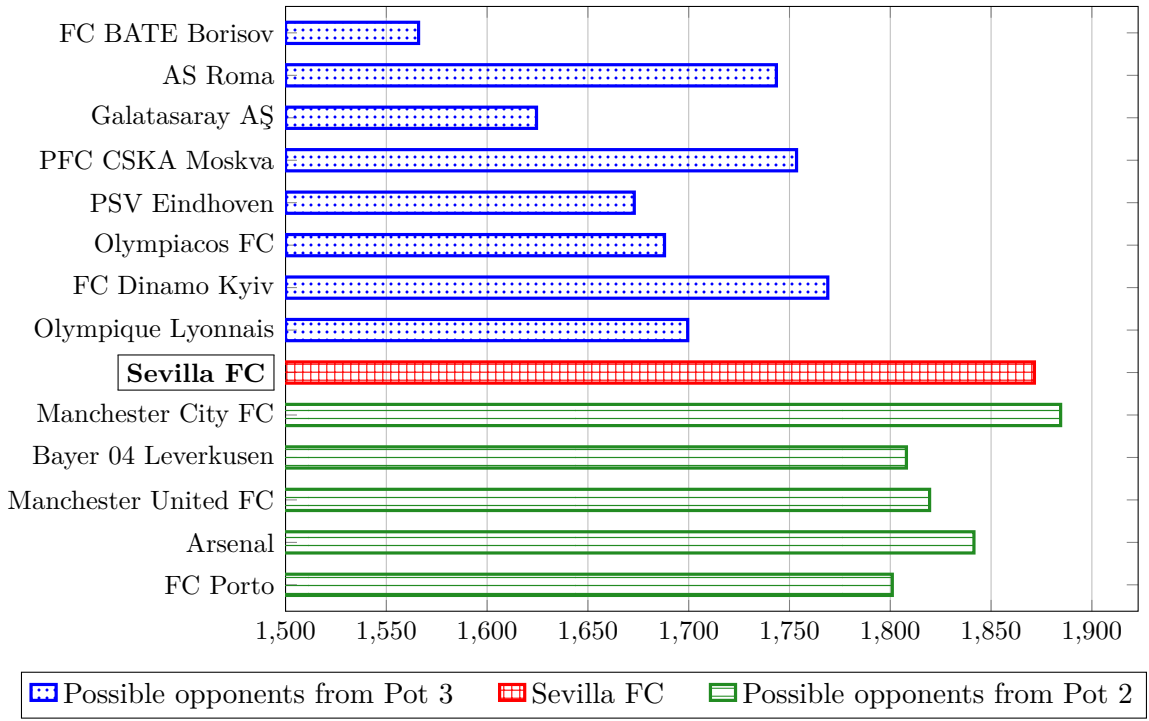
\begin{figure}[!ht]
\centering
\caption{The average Elo rating of the possible opponents of Sevilla FC in the group stage of the 2015/16 UEFA Champions League}
\label{Fig1}

\begin{tikzpicture}
%\selectcolormodel{gray}
\begin{axis}[width = 0.8\textwidth, 
height = 0.6\textwidth,
xmin = 1500,
xmajorgrids,
%ymajorgrids,
%xlabel = Elo rating,
ylabel style = {font = \small},
symbolic y coords = {FC Porto, Arsenal, Manchester United FC, Bayer 04 Leverkusen, Manchester City FC, Sevilla FC, Olympique Lyonnais, FC Dinamo Kyiv, Olympiacos FC, PSV Eindhoven, PFC CSKA Moskva, Galatasaray A\c{S}, AS Roma, FC BATE Borisov},
ytick = data,
yticklabels = {FC Porto, Arsenal, Manchester United FC, Bayer 04 Leverkusen, Manchester City FC, \framebox{\textbf{Sevilla FC}}, Olympique Lyonnais, FC Dinamo Kyiv, Olympiacos FC, PSV Eindhoven, PFC CSKA Moskva, Galatasaray A\c{S}, AS Roma, FC BATE Borisov},
%x tick label style={rotate=90,anchor=east},
enlarge y limits = {abs = 0.35cm},
xbar stacked,
bar width = 8pt,
legend entries={Possible opponents from Pot 3 $\quad$, Sevilla FC $\quad$, Possible opponents from Pot 2},
legend style = {at = {(0.35,-0.1)},anchor = north,legend columns = 4,font = \small}
]
%\draw (axis cs:\pgfkeysvalueof{Switzerland,0)  -- (axis cs:\pgfkeysvalueof{Kosovo,0);
\addplot[blue, pattern color = blue, pattern = dots, very thick] coordinates {
(0,FC Porto)
(0,Arsenal)
(0,Manchester United FC)
(0,Bayer 04 Leverkusen)
(0,Manchester City FC)
(0,Sevilla FC)
(1699.5,Olympique Lyonnais)
(1769,FC Dinamo Kyiv)
(1688,Olympiacos FC)
(1673,PSV Eindhoven)
(1753.5,PFC CSKA Moskva)
(1624.5,Galatasaray A\c{S})
(1743.5,AS Roma)
(1566,FC BATE Borisov)
};

\addplot[red, pattern color = red, pattern = grid, very thick] coordinates {
(0,FC Porto)
(0,Arsenal)
(0,Manchester United FC)
(0,Bayer 04 Leverkusen)
(0,Manchester City FC)
(1871.5,Sevilla FC)
(0,Olympique Lyonnais)
(0,FC Dinamo Kyiv)
(0,Olympiacos FC)
(0,PSV Eindhoven)
(0,PFC CSKA Moskva)
(0,Galatasaray A\c{S})
(0,AS Roma)
(0,FC BATE Borisov)
};

\addplot[ForestGreen, pattern color = ForestGreen, pattern = horizontal lines, very thick] coordinates {
(1801,FC Porto)
(1841.5,Arsenal)
(1819.5,Manchester United FC)
(1808,Bayer 04 Leverkusen)
(1884.5,Manchester City FC)
(0,Sevilla FC)
(0,Olympique Lyonnais)
(0,FC Dinamo Kyiv)
(0,Olympiacos FC)
(0,PSV Eindhoven)
(0,PFC CSKA Moskva)
(0,Galatasaray A\c{S})
(0,AS Roma)
(0,FC BATE Borisov)
};
\end{axis}
\end{tikzpicture}
\end{figure}

%\end{document}

Figure~\ref{Fig1} highlights that Sevilla FC is remarkably better off in our hypothetical scenario if it would be drawn from Pot 2 as all of the eight possible opponents are weaker than any possible opponent if the club would be drawn from Pot 3.
%due to its better performance in its domestic championship.
The expected Elo rating of the eight teams from Pot 3 is $1689.625$, while this value is $1830.9$ for the five teams from Pot 2. Since Sevilla FC has an Elo rating of $1871.5$, it wins with a 74.02\% probability against the former teams, but only with a 55.82\% probability against the latter clubs according to the standard formula $1/ \left( 1 + 10^{- \Delta/400} \right)$, where $\Delta$ is the difference between the Elo ratings of the two teams. This is a robust difference, especially because it can be crucial with respect to qualifying for the knockout stage as only the first two teams from each group advance, while the third is relegated to the Round of 32 in the Europa League.

On the other hand, the average Elo rating of possible opponents from Pot 1 (Chelsea FC, FC Bayer M\"unchen, Juventus, SL Benfica, Paris Saint-Germain FC, FC Zenit St Petersburg) is $1864.67$ if Sevilla FC is drawn from Pot 2, but is reduced to $1837.29$ because PSV Eindhoven enters Pot 1 if Sevilla FC is drawn from Pot 3.

While this secondary effect somewhat mitigates the problem of perverse incentives, it remains clear that Sevilla FC loses due to its better performance against another Spanish team. Using a more sophisticated quantification method may give a better estimation on the size of this negative effect, however, without changing our main finding:
the ill-designed seeding regime from the 2015/16 season of the UEFA Champions League can severely harm an innocent team merely for scoring more points in its domestic championship.

The Champions League is regulated in three-year cycles, thus the scenario outlined in Section~\ref{Sec2} could have emerged in the three seasons played between 2015 and 2018.
Since the 2018/19 season, the titleholder of the Europa League from the previous year automatically qualifies for the group stage of the Champions League, too, where it is seeded in Pot 1. Consequently, the top pot consists of the two titleholders and the champions of the six highest-ranked associations. Furthermore, all vacancies are awarded to the champion(s) of the next highest-ranked association(s) as \citet[Article~13.06]{UEFA2019a} describes for the 2019/20 season. It means that the problem has probably become worse because of its possible occurrence in two national leagues, although there were no such vacancies in the 2018/19 and 2019/20 seasons.

The policy of guaranteeing a place in the top pot to certain champions can also be criticised for creating unbalanced groups \citep[Section~3.4]{Guyon2019b}.
For example, FC Lokomotiv Moskva was in Pot 1 in the 2018/19 season as the champion of Russia (the sixth-ranked UEFA association), while its UEFA club coefficient would have placed the team only in Pot 4. Unsurprisingly, FC Lokomotiv Moskva finished fourth in its group, and two ``lucky'' teams, the Portuguese FC Porto from Pot 2 and the German FC Schalke 04 from Pot 3 had an easy path to the Round of 16.
In addition, forming Pot 1 on the basis of national leagues and ignoring this principle for the other pots is inconsistent, difficult to justify, and unfair to the champion of the next best league \citep{Guyon2015b}.
Interestingly, there is no such differentiation in the UEFA Europa League, where all pots are created based on the UEFA club coefficients.

\section{Policy implications} \label{Sec4}

Rewarding league champions in a tournament called Champions League seems to be a reasonable principle, even though it is moving farther from its original concept of being a ``league of champions'' \citep{Csato2020h}.
However, the current definition of Pot 1 remains unfair. It is a shame because there exists a straightforward solution, revealed by \citet[Proposition~3]{DagaevSonin2018}: all vacancies should be filled through the round-robin tournament, i.e.\ the national leagues. That is, \citet[Article~13.06]{UEFA2019a} should be modified in the following way: \\
``\emph{For the purpose of the draw, the 32 clubs involved in the group stage are seeded into four groups of eight. The first group comprises the titleholder (top seed), the UEFA Europa League titleholder and the domestic champions of the six associations ranked highest in the access list (see Annex A). If either or both titleholders are the domestic champions of one of the top six associations, the group is completed with \textbf{the runner(s)-up (and the third-placed club) of the same association(s)}. The other three groups are composed in accordance with the club coefficient rankings established at the beginning of the season (see Annex D).}''

This proposal immediately guarantees incentive compatibility because no champion can gain a slot in Pot 1 due to the identity of the titleholder(s), hence no team would be relegated to a lower pot merely by having better results in its domestic league. With this policy, Sevilla FC would have been placed into Pot 2 regardless of which team would have won La Liga in our hypothetical example of Section~\ref{Sec2}. UEFA is encouraged to introduce the suggested amendment in the Champions League from the 2021-24 cycle onwards.

Naturally, there are further policies that can remedy the problem of incentive incompatibility. The pre-2015 seeding regime formed the pots based on the UEFA club coefficients, except for automatically placing the titleholder in the first pot. This guarantees strategy-proofness. \citet{Guyon2015b} recommends a fundamental reform of the seeding, which does not suffer from the lack of win incentive, and solves further fairness issues, too.
However, the above modification remains the minimal one that eliminates misaligned incentives.

\section{Conclusions} \label{Sec5}

Regulations that allow for a successful tanking strategy or a punishment of a team when it scores more points threaten the integrity of sports and are against the spirit of the game. 
Therefore, the design of a sports tournament remains an important field of analysis for game theory and operations research. In our opinion, the scientific community has a responsibility to present all possible cases of incentive incompatibility, regardless of the frequency of dubious situations.

We have revealed that the seeding of the clubs into pots in the group stage of the UEFA Champions League, the most prestigious annual club football competition in Europe, suffers from perverse incentives since the 2015/16 season because vacancies in the top pot are filled through an ill-constructed policy.
Hopefully, this work will persuade the decision makers to implement our straightforward proposal for solving the problem before it causes controversy.

\section*{Acknowledgements}
\addcontentsline{toc}{section}{Acknowledgements}
\noindent
Discussions with \emph{Julien Guyon} gave an important motivation for the paper. \\
Four anonymous reviewers provided valuable comments and suggestions on earlier drafts. \\
% \emph{Tam\'as Halm}
We are indebted to the \href{https://en.wikipedia.org/wiki/Wikipedia_community}{Wikipedia community} for contributing to our work by collecting and structuring useful information on the tournaments discussed. \\
This work was supported by the MTA Premium Postdoctoral Research Program grant PPD2019-9/2019.

\bibliographystyle{apalike}
\bibliography{All_references}

\end{document}